# VERTICAL PROJECTILE MOTION TAKING INTO ACCOUNT THE WIND

P. Chudinov

Department of Engineering, Perm State Agro-Technological University, 614990, Perm, Russia
chupet@mail.ru

**Abstract**

We consider the problem of the motion of a projectile thrown vertically upward from a surface. In addition to gravity, the drag force of the medium is taken into account, which is considered a quadratic function of the relative velocity of the projectile. It is also assumed that the projectile is acted upon by a wind that has a constant vertical velocity in magnitude and direction. Analytical solutions of differential equations of motion, including the influence of wind, are written down. As an example of the application of the obtained formulas, the movement of a badminton shuttlecock is considered. Various features of movement are considered, such as the shuttlecock "hovering" in the air and continuous upward movement. The educational and methodological significance of the obtained results for students of various levels of training is noted.

## INTRODUCTION

The problem of the motion of a projectile thrown at an angle to the horizon is a classic one and has a rich history. An important special case of this problem is the study of vertical motion of a projectile. The scientific study of such motion began with Galileo Galilei (Galilei, 1638) and continues to the present day (Lindemuth, 1971; Vial, 2012; Mungan & Rittenhouse, 2021). Unlike other works, this work studies the movement in the air of a projectile thrown vertically upward, taking into account the wind. The wind speed is assumed to be constant and directed along the positive direction of the vertical *y*-axis. The resistance of the medium is assumed to be quadratic in speed. In this study the conditions of applicability of the quadratic resistance law are deemed to be fulfilled, i.e. Reynolds number *Re* lies within $1×10^3 < Re < 2×10^5$. In this formulation, the problem is, in a certain sense, a model one, since it is difficult to implement the throwing conditions with such a wind direction. Nevertheless, such a task is methodologically useful.

The problem statement is as follows. A projectile of mass *m* is located on the surface at the initial point *O* of the vertical axis *y*. At the instant of time, the projectile is given an initial velocity $V_0$ in the vertical direction. When moving, the projectile is affected by the force of gravity, the force of air resistance and of the wind, which is constant in magnitude and direction. It is required to study the motion of the projectile under given conditions. Figure 1 shows the motion of the projectile. The left figure corresponds to the upward movement of the projectile **M** relative to the surface, the right figure corresponds to the downward movement. Here **V** is the projectile absolute velocity vector, **P** = *m***g** is the gravity vector, **R** is the vector of the environmental resistance force, $\mathbf{W_y}$ is the wind speed vector.

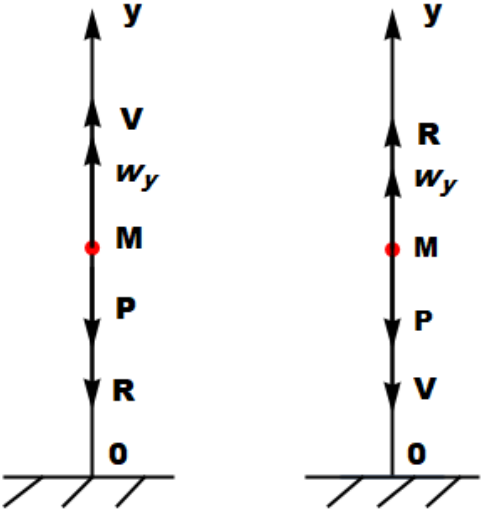

**Figure 1.** Forces acting on the projectile and vectors of its speed and wind speed.

## EQUATIONS OF PROJECTILE MOTION AND METHOD OF THEIR SOLUTIONS

The equation of projectile motion in the vector form is

$$m\frac{d\mathbf{V}}{dt} = m\mathbf{g} + \mathbf{R}.$$

The drag force of the medium is assumed to be a quadratic function of velocity. In the presence of wind, the drag force of the medium in the quadratic drag model has the form (Lubarda&Lubarda, 2022)

$$\mathbf{R} = -c|\mathbf{V} - \mathbf{w_y}|(\mathbf{V} - \mathbf{w_y}).$$

Here $c$ is the resistance coefficient. For the convenience of further calculations, we present the drag coefficient in the form $c = mgk$. The introduced coefficient $k$ can be conveniently determined through the terminal velocity of the projectile $V_{term}$ (Cohen et al, 2014): $k = 1/V_{term}^2$. Terminal velocity is the maximum velocity attainable by an object as it falls through a fluid (air is the most common example). Then the equations of vertical motion of the projectile in projection onto the y-axis will take the form

$$\frac{dy}{dt} = V, \quad \frac{dV}{dt} = -g \mp gk(V - w_y)^2. \quad (1)$$

The minus sign in equation (1) corresponds to the upward movement of the projectile when the condition is met $w_y \leq V \leq V_0$. The plus sign corresponds to the downward movement of the projectile or up when the condition is met $V \leq w_y$.

First, we consider the upward movement of the object. Initial motion conditions:

$$y(t=0) = y_0 = 0, \quad V(t=0) = V_0, \quad w_y = const.$$

Equations of motion (1) admit analytical solutions described by elementary functions. Omitting intermediate calculations, we write down the dependencies $V(t), y(t), y(V)$ for the case of upward movement when the condition is met $w_y \leq V \leq V_0$:

$$V(t) = \frac{1}{\sqrt{k}} \tan\left[\arctan\left(\sqrt{k}(V_0 - w_y)\right) - g\sqrt{k}t\right] + w_y,$$

$$y(t) = \frac{1}{gk} \ln\left[\frac{\cos\left(\arctan\left(\sqrt{k}(V_0 - w_y)\right) - g\sqrt{k}t\right)}{\cos\left(\arctan\left(\sqrt{k}(V_0 - w_y)\right)\right)}\right] + w_y t,$$

$$y(V) = \frac{w_y}{g\sqrt{k}}\left[\arctan\left(\sqrt{k}(w_y - V)\right) - \arctan\left(\sqrt{k}(w_y - V_0)\right)\right]$$

$$+ \frac{1}{2gk} \ln\left[\frac{1 + k(V_0 - w_y)^2}{1 + k(V - w_y)^2}\right]. \quad (2)$$

Using formulas (2) we find the values of $t_1$ and $H_1$, corresponding to the condition $V = w_y$:

$$t_1 = \frac{1}{g\sqrt{k}} \arctan\left[\sqrt{k}(V_0 - w_y)\right], \quad H_1 = y(t_1).$$

Now let's write down similar formulas for $V(t), y(t), y(V)$, describing the downward movement of the object:

$$V(t) = w_y + \frac{1}{\sqrt{k}} \tanh\left[g\sqrt{k}(t_1 - t)\right],$$

$$y(t) = H_1 + w_y(t - t_1) - \frac{1}{gk} \ln\left[\cosh\left(g\sqrt{k}(t - t_1)\right)\right],$$

$$y(V) = H_1 + \frac{1}{2gk} \ln\left[1 - k(V - w_y)^2\right] \quad (3)$$

$$- \frac{w_y}{2g\sqrt{k}} \operatorname{arctanh}\left[\sqrt{k}(V - w_y)\right].$$

When moving down, the initial conditions of movement are as follows:

$$t = t_1, \quad y(t_1) = H_1, \quad V(t_1) = w_y.$$

Equations (2) – (3) generalize the known equations of vertical motion with resistance by taking into account the influence of wind. The value of the resistance coefficient $k = 0$ cannot be used in these formulas, since division by zero occurs.

However, at very small values of the parameter $k$ ($k = 10^{-12}$) and in the absence of wind, formulas (2) – (3) transform into the well-known formulas of the classical theory of projectile motion without air resistance.

This statement is illustrated by Figure 2. The dependence $y = y(V)$ for different throwing conditions is constructed using the same formulas (2) – (3). Parabola 1

$$y(V) = \frac{V_0^2 - V^2}{2g}$$

describes movement without resistance and without wind, curves 2, 3 describe movement with both the quadratic resistance of the medium and of the wind. The uniqueness of curves 2, 3 will be explained below.

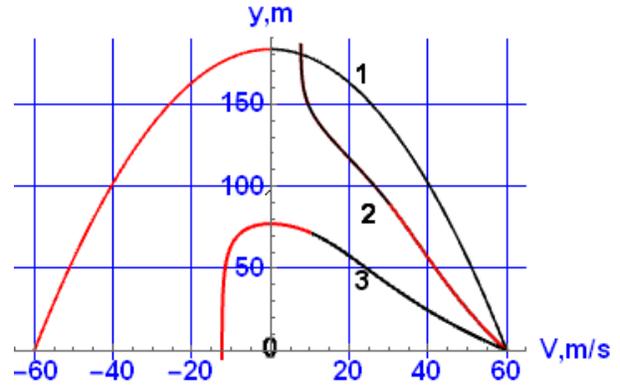

**Figure 2.** Dependency graphs $y(V)$. Throwing conditions for curves 1, 2, 3 respectively:

1 – $V_0 = 60$ m/s, $k = 10^{-12}$ s$^2$/m$^2$, $w_y = 0$;

2 – $V_0 = 60$ m/s, $k = 0.002$ s$^2$/m$^2$, $w_y = 30$ m/s;

3 – $V_0 = 60$ m/s, $k = 0.002$ s$^2$/m$^2$, $w_y = 10$ m/s.

## CALCULATION RESULTS AND ANALYSIS

As an example of using the obtained formulas, we consider the movement of a badminton shuttlecock. This sports projectile is remarkable because it has the lowest terminal speed among other sports projectiles (tennis ball, baseball, golf ball, etc.) $V_{term} = 6.75$ m/s.

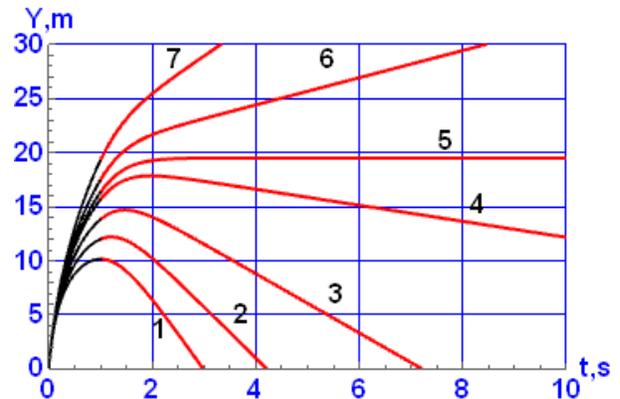

**Figure 3.** Graphs of the function $y = y(t)$.

Throwing conditions $V_0 = 60$ m/s, $k = 0.022$ s$^2$/m$^2$.

The used wind speed values are given in Table 1. Functional dependencies of the problem are plotted in Figures 3, 4, 5 – accordingly $y(t), V(t), y(V)$.

**Table 1.** Curve numbers and corresponding wind speed values.

| Curve № of Figure 3 | 1 | 2 | 3 | 4 | 5 | 6 | 7 |
|---|---|---|---|---|---|---|---|
| Curve № of Figure 4 | 1 | 2 | | | 3 | 4 | 5 |
| Curve № of Figure 5 | 1 | 2 | | | 3 | 4 | 5 |
| $w_y$, m/s | 0 | 2 | 4 | 6 | 6.75 | 8 | 10 |

Curves 1 – 7 in Figure 3 are plotted for the corresponding values of wind speed given in the last line of Table 1. Curves 1–4 are plotted under the condition $w_y < V_{term}$. The object reaches the top of the trajectory, then returns to the surface. For curve 5, equality holds $w_y = V_{term}$. Under this condition, the projectile reaches the top of the trajectory and remains at this constant height with the further passage of time. Let's call this trajectory "hovering". Curves 6 and 7 are plotted under the condition $w_y > V_{term}$. The y coordinate continuously increases, the shuttlecock rises vertically upward under the influence of the wind. Black in Figure 3 shows sections of the dependencies $y(t)$, corresponding to the movement of the projectile from the surface to the trajectory point $y(t_1)$. The speed of the projectile $V$ is within the limits $V_0 \geq V \geq w_y$. The sections of dependencies $y(t)$, corresponding to the further movement of the projectile are shown in red. For curves 1 – 4 this movement represents a vertical fall onto the surface; for curves 6 and 7 it is a further upward movement vertically at a constant speed $V = w_y - V_{term}$ (see Figure 5). For curve 5, after reaching the top, the coordinate remains constant: $y = H$. When moving along curves 5, 6, 7, the projectile remains in the air for an unlimited time.

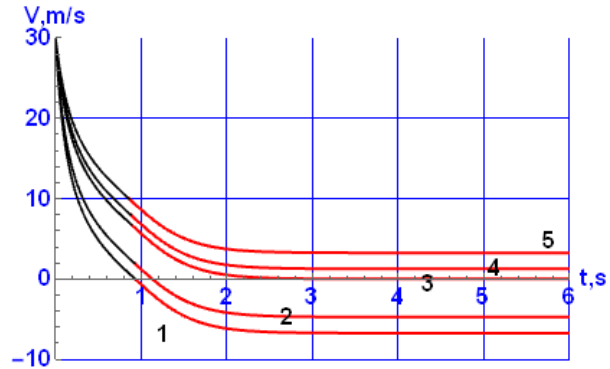

**Figure 4.** Graphs of the function $V = V(t)$.

Throwing conditions: $V_0 = 30$ m/s, $k = 0.022 \text{ s}^2/\text{m}^2$.

Figure 4 shows graphs of the function $V(t)$. Under the condition $w_y > V_{term}$, the projection of the shuttlecock speed onto the y-axis satisfies the inequality $V_y \geq 0$. In both cases equality holds $V_y = w_y - V_{term}$. For curve 2 the final speed $V = 2 - 6.75 = -4.75$ m/s, for curve 5 the speed of steady motion is $V = 10 - 6.75 = 3.25$ m/s.

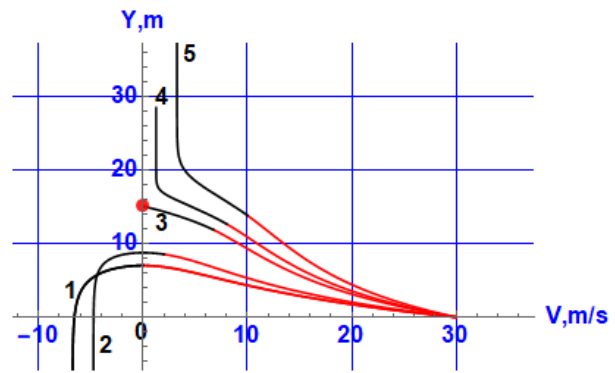

**Figure 5.** Graphs of the function $y(V)$.

Throwing conditions: $V_0 = 30$ m/s, $k = 0.022 \text{ s}^2/\text{m}^2$.

In Figure 5 the dependencies $y(V)$ are plotted. Curves 1, 2, 3, 4, 5 are respectively plotted at wind speed values $w_y =$ 0; 2; 6.75; 8; 10 m/s. Curve 3 characterizes the "hanging" trajectory. The absolute speed of the object at the top of the trajectory is zero, the height is constant. The object is motionless relative to the surface and is represented by a red dot. For curve 1, the modulus of velocity at the moment of falling onto the surface is almost equal to $V_{term}$, for curve 2 the modulus of velocity at the moment of falling to the ground is equal $V = V_{term} - w_y$. The projection of the velocity vector onto the *y*-axis is negative for curves 1 and 2. For curves 4 and 5, the speed of the shuttlecock rising vertically upward is positive and equal to $V = w_y - V_{term}$. Summarizing, it should be noted that both when moving down (curves 1, 2) and when moving up (curves 4, 5), the projection of the shuttlecock speed on the y-axis along with its sign is equal to $V_y = w_y - V_{term}$. The meaning of the black and red colors of the curve sections $y(V)$ is similar to that in Figures 3, 4.

The value of wind speed $w_y = V_{term}$ divides the nature of projectile movement into two types. When $w_y < V_{term}$ the projectile first moves up, then down. When set to $w_y > V_{term}$ value, the projectile moves only upward. When set to $w_y = V_{term}$ value, the projectile hangs in the air motionless relative to the surface. It should be noted that the black sections of the curves in Figures 3 and 4 were constructed using formulas (2). The sections of the red curves are constructed using formulas (3)), which are intended to study the downward movement of the projectile. The movement of a projectile upward with speed $V = w_y - V_{term}$ on curves 6 and 7 of Figure 3 and curves 4 and 5 of Figure 5 can be interpreted as "falling down" with speed $V = V_{term}$ in a flow of a medium moving upward with speed $V = w_y$. Therefore, this movement is described by formulas (3).

Time values $t_a$ and coordinates $H_a$ at the trajectory tops in Fig. 3 are determined by the formulas

$$t_a = t_1 + \frac{1}{g\sqrt{k}} \operatorname{arctanh}\left(\sqrt{k} w_y\right),$$

$$H_a = H_1 + w_y(t_a - t_1) - \frac{1}{gk} \ln\left[\cosh\left(g\sqrt{k}(t_a - t_1)\right)\right].$$

## CONCLUSION

The main conclusion of the study is that the introduction of the influence of wind into the equations of vertical motion of a projectile adds two important features to the general nature of the motion (up - down). The first feature is that when the condition $w_y = V_{term}$ is met, the projectile "hangs" in the air indefinitely at the top of the trajectory. The second feature is that the nature of the movement only up, subject to $w_y > V_{term}$. This movement is realized when the wind speed is greater than the terminal speed of the projectile. We emphasize that the work considers a model problem, but useful in educational and methodological terms. The classical problem of the vertical movement of a projectile in a uniform field of gravity can be successively supplemented (depending on the level of training of students) by introducing the drag force of the environment and the influence of wind.

## ACKNOWLEDGMENTS

The authors thank Dr. D.V. Zitta for useful advice.

## REFERENCES

Christensen, R., Teiwes, R., Petersen, S., Uggerhøj, U., &Jacoby, B. (2014). Laboratory test of the Galilean universality of the free fall experiment. *Physics Education.* 49(2), 201–210. https://doi.org/10.1088/0031-9120/49/2/201

Chudinov, P., Eltyshev, V., & Barykin, Y. (2022). Projectile motion in midair using simple analytical approximations. *The Physics Teacher.* 60(9), 774–778. https://doi.org/10.1119/5.0053162

Cohen, C. et al. (2014). The aerodynamic wall. *Proceedings of the Royal Society A.* 470, 20130497. HTTPS://DOI.ORG/10.1098/RSPA.2013.0497

Galilei, G. (1638). *Dialogues Concerning Two New Sciences.* Translated by H. Crew and A. de Salvio from original edition 1638. Dover Publications, New York, 1914. http://books.google.com/books?id=SPhnaiERbWcC&oe=UTF-8

Lindemuth, J. (1971). The effect of air resistance on falling balls, *American Jounal of Physics,* 39(7), 757–759. https:/DOI:10.1119/1.1986278

Lubarda, M. & Lubarda, V. (2022). A review of the analysis of wind-influenced projectile motion in the presence of linear and nonlinear drag force. *Archive of Applied Mechanics.* 92, 1997–2017. https://doi.org/10.1007/s00419-022-02173-7

Mungan, C. (2006). Energy-based solution for the speed of a ball moving vertically with drag. *European Journal of Physics,* 27(5) 1141. https:/DOI 10.1088/0143-0807/27/5/013

Mungan, C., Rittenhouse, S., & Lipscombe, T. (2021). A vertical race up and back down with and without drag. *American Journal of Physics,* 89(1), 67–71. https://doi.org/10.1119/10.0001893

Soares, A., Caramelo, L., & Andrade, M. (2012). Study of the motion of a vertically falling sphere in a viscous fluid. *European Journal of Physics.* 33(5), 1053–1062. https://doi.org/10.1088/0143-0807/33/5/1053

Timmerman, P., & van der Weele, J. (1999). On the rise and fall of a ball with linear or quadratic drag. *American Jounal of Physics,* 67(6), 538–546. https://pubs.aip.org/aapt/ajp/article-abstract/67/6/538/1055298/

Vial, A. (2012). Fall with linear drag and Wien's displacement law: Approximate solution and Lambert function. *European Journal of Physics.* 33(4), 751–755. https://doi.org/10.1088/0143-0807/33/4/751